\documentclass[aps,prb,twocolumn,superscriptaddress,showpacs]{revtex4}

\usepackage{amsmath}
\usepackage[dvips,dvipdfm]{graphicx}
\usepackage[usenames,dvipsnames]{color}

\begin{document}
\title{Effect of an InP/In$_{0.53}$Ga$_{0.47}$As Interface on Spin-orbit Interaction in In$_{0.52}$Al$_{0.48}$As/In$_{0.53}$Ga$_{0.47}$As Heterostructures}
\author{Yiping Lin}
\altaffiliation{Present address: Department of Physics, National Tsing Hua University, Hsinchu 300, Taiwan.}
\affiliation{NTT Basic Research Laboratories, NTT Corporation, Atsugi, Kanagawa 243-0198, Japan}
\author{Takaaki Koga}
\altaffiliation{Present address: Graduate School of Information Science and Technology, Hokkaido University, Sapporo 060-0814, Japan.}
\affiliation{NTT Basic Research Laboratories, NTT Corporation, Atsugi, Kanagawa 243-0198, Japan}
\affiliation{PRESTO, Japan Science and Technology Agency, Kawaguchi, Saitama 332-0012, Japan}
\author{Junsaku Nitta}
\affiliation{NTT Basic Research Laboratories, NTT Corporation, Atsugi, Kanagawa 243-0198, Japan}
\affiliation{CREST, Japan Science and Technology Agency, Kawaguchi, Saitama 332-0012, Japan}
\date{\today}

\begin{abstract}
We report the effect of the insertion of an InP/In$_{0.53}$Ga$_{47}$As Interface on Rashba spin-orbit interaction in In$_{0.52}$Al$_{0.48}$As/In$_{0.53}$Ga$_{0.47}$As quantum wells. A small spin split-off energy in InP produces a very intriguing band lineup in the valence bands in this system. With or without this InP layer above the In$_{0.53}$Ga$_{47}$As well, the overall values of the spin-orbit coupling constant $\alpha$ turned out to be enhanced or diminished for samples with the front- or back-doping position, respectively. These experimental results, using weak antilocalization analysis, are compared with the results of the $\mathbf{k\cdot p}$ theory. The actual conditions of the interfaces and materials should account for the quantitative difference in magnitude between the measurements and calculations.
\end{abstract}
\maketitle
Spin-orbit (SO) interaction provides a central mechanism for the realization of optical spin orientation and detection, and, in general, is responsible for spin relaxation. This relaxation causes the spin of an electron to precess during the time of flight. Utilizing this interaction, several applications have been proposed, both in the ballistic region~\cite{Datta90,Nitta99} and diffusive region,~\cite{Schliemann03,Cartoixa03} as spin field effect transistors or spin inferometers. Inspired by these proposals, it is essential for us to investigate the ways of manipulating electron spins using the SO coupling.

The mechanisms for the SO interaction in semiconductors can be categorized into the Dresselhaus~\cite{Dresselhaus55} and Rashba terms.~\cite{Rashba60,Bychkov84} The former originates from the bulk inversion asymmetry (BIA), a characteristic of zincblende semiconductors, and the latter comes from the structural inversion asymmetry (SIA). Their relative strength depends on the choice of materials.~\cite{Lommer88}
 In the system of concern here, \emph{i.e.} an In$_{0.53}$Ga$_{0.47}$As quantum well (QW), SIA is frequently considered as the main contribution to the SO interaction.~\cite{Das90,Nitta97,Engles97,Koga02b} For the Rashba term in the SO interaction, a counter-intuitive fact is that it is the valence-band structure that determines its coupling constant (\emph{not} the conduction-band profile) in the $\mathbf{k\cdot p}$ theory [see Eq.~(\ref{eq:a})]. In this respect, it is of fundamental interest to study the SO coupling constant including the details of valence-band alignment, which highlights the \emph{interface effect}.

In transport measurements, it is common to determine the SO coupling constant from the beating pattern in Shubnikov--de Haas (SdH) oscillations.~\cite{Das89,Das90,Nitta97,Engles97} However, the absence of beating nodes does not exclude the existence of the SO interaction.~\cite{Koga02b} It was suggested that the trace of SO interaction in high-mobility GaAs samples can be revealed by applying microwave excitation with varying frequencies.~\cite{Mani04} Alternatively, the SO coupling constant can be extracted from the analysis of weak antilocalization (WAL).~\cite{Dresselhaus92,Chen93,Knap96,Kreshchuk98,Koga02b,Studenikin03b} This method works especially well for samples with low mobilities and strong SO interactions: for the former, in many cases the fields at which SdH oscillations start to be visible are so high that the beating nodes cannot be observed; for the latter, the required frequency in photoexcitation is hard to achieve. In this paper, we study the interface effect of the SO coupling constant from the WAL measurements.
\begin{figure}[!ht]
\includegraphics[width=2.3in,clip=]{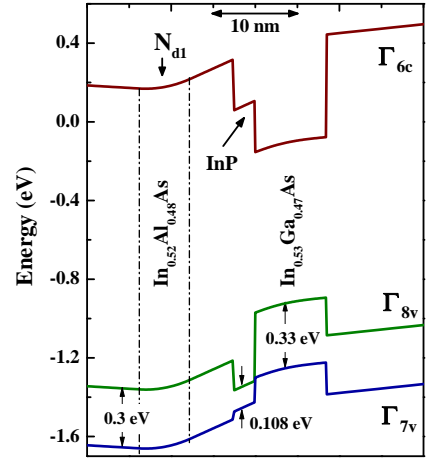}
\caption{\label{fig:bands}Band-structure profile of No. 1 obtained through the self-consistent calculation of Poisson and Schr\"odinger equations at $\Gamma$ point of the Brillouin zone. $\Gamma_{6c}$, and $\Gamma_{8v}$ and $\Gamma_{7v}$ are the conduction band and valence bands, respectively. The indicated energies are the spin split-off energies. N$_{d1}$ is the doping concentration above the QW.}
\end{figure}

Materials like In$_{x}$Al$_{1-x}$As, In$_{x}$Ga$_{1-x}$As, and InP have been studied extensively and considered to be useful in many device applications. Since InP has a relatively small spin split-off energy ($\Delta_{SO}$) in this material family, InP can be a good candidate for studying the interface effect from the point of view of valence bands.~\cite{Schapers98} For a lattice-matched system, the valence band ($\Gamma_{8v}$) of InP is lower than the split-off band ($\Gamma_{7v}$) of In$_{0.53}$Ga$_{0.47}$As in energy, as shown in Fig.~\ref{fig:bands}. Therefore, inserting an InP layer between In$_{0.52}$Al$_{0.48}$As and In$_{0.53}$Ga$_{0.47}$As provides an unique band alignment for $\Gamma_{7v}$ and $\Gamma_{8v}$ bands at the interface. In combination with the interface effect, the doping position with respect to the QW can modify the band bending in the QW and thereby, the gate-voltage dependence of the SO interaction. There have been some works on the SO interaction using InP in sample design.~\cite{Schapers98,Kreshchuk98,Studenikin03b} The present work differs from them in that our focus is on how the SO interaction is modified by the combination of the interface effect and the doping position.

Four samples of In$_{0.52}$Al$_{0.48}$As/(InP/)In$_{0.53}$Ga$_{0.47}$As QWs were grown on the InP substrates by metalorganic chemical vapor deposition. Two samples, one with and one without an InP layer at the top In$_{0.52}$Al$_{0.48}$As/In$_{0.53}$Ga$_{0.47}$As interface, had a doping layer above the QW (No. 1 and No. 3, respectively), while the other two, one with and one without an InP layer, had a doping layer below the QW (No. 2 and No. 4, respectively). The layer structures of these samples are listed in Table~\ref{table}. 
The $n$-type doping concentration (Si) and the thickness of In$_{0.53}$Ga$_{0.47}$As QW were designed such that the samples had similar carrier densities ($N_{S}$) for the two-dimensional electron gases (2DEGs) at zero gate voltage. Samples were fabricated using the conventional photolithographic technique with 1000 {\AA} Au as front gate. Measurements were carried out in a $^{3}$He cryostat (0.3K) with magnetic fields applied perpendicular to the sample surface.
\begin{table}
\caption{\label{table}Active layer structures of four samples, which is listed from the sample surface to the buffer layer (before InP substrate). Gate (not listed) is on the top. Thickness in $\textrm{\AA}$.~\cite{TEM}}
\vspace{0.2cm}
\begin{ruledtabular}
\begin{tabular}{ccccc}
& No. 1& No. 2 & No. 3 & No. 4\\
\hline
In$_{0.52}$Al$_{0.48}$As & 250 & 360 & 250 & 370\\
n-In$_{0.52}$Al$_{0.48}$As\footnote{N$_{d1}$=2.5$\times 10^{18}$ cm$^{-3}$} & 60 & -- & 60 & --\\
In$_{0.52}$Al$_{0.48}$As & 50 & -- & 60 & --\\
InP & 25 & 25 & -- & --\\
In$_{0.53}$Ga$_{0.47}$As & 85 & 85 & 100 & 100 \\
In$_{0.52}$Al$_{0.48}$As & -- & 60 & -- & 60\\
n-In$_{0.52}$Al$_{0.48}$As\footnote{N$_{d2}$=2$\times 10^{18}$ cm$^{-3}$}  & -- & 60 & -- & 60\\
In$_{0.52}$Al$_{0.48}$As & 2120 & 2000& 2120 & 2000\\
\end{tabular}
\end{ruledtabular}
\end{table}

The Hamiltonian for the Rashba term is written as~\cite{Bychkov84}
\begin{equation}
\mathcal{H}_{so} = \alpha (\sigma_{x}k_{y}-\sigma_{y}k_{x}) = \boldsymbol{\sigma} \cdot \boldsymbol{\Omega_{1}},
\end{equation}
where $\alpha$ is the Rashba spin-orbit coupling constant. $\boldsymbol{\sigma}=(\sigma_{x},\sigma_{y})$ and $\boldsymbol{\Omega_{1}}= (\Omega_{1}^{R} \sin \psi, -\Omega_{1}^{R} \cos \psi)$ are 2D vectors in the plane of QW, where $\Omega_{1}^{R}=\alpha k$ and $\tan \psi = k_{y}/k_{x}$. We used the model developed by Iordanskii \emph{et al.}~\cite{Iordanskii94} for the conductivity correction $\Delta \sigma(H)$, where $H$ is the magnetic field, in which only the D'yakonov-Perel' is responsible for the spin relaxation. The only two adjustable parameters in fitting the experimental data  with this model are: (i) $H_{\varphi}$, the magnetic field related to the phase coherent relaxation time $\tau_{\varphi}$ and (ii) $H_{SO}$, the magnetic field related to the spin splitting energy. When only the Rashba term is present:
\begin{equation}\label{eq:Hso}
H_{\varphi}=\frac{\hbar}{4De\tau_{\varphi}}\ \textrm{and}\ 
H_{so} = \frac{\hbar}{4De} \frac{{2(\Omega_{1}^{R})}^{2} \tau_{tr}}{\hbar^{2}}.
\end{equation}
Here $D$ is the diffusion constant and $\tau_{tr}$ is momentum relaxation time. These parameters were obtained from the results of Hall and SdH measurements. The extracted $\alpha$ values were then compared with the calculated ones using the $\mathbf{k \cdot p}$ formalism:~\cite{Schapers98}
\begin{equation}\label{eq:a}
\alpha = \frac{\hbar^{2}E_{p}}{6m_{0}} \big\langle \Psi \big | \frac{d}{dz} \big( \frac{1}{E_{F}-E_{\Gamma_{7}}(z)} - \frac{1}{E_{F} - E_{\Gamma_{8}}(z)} \big) \big | \Psi \big \rangle,
\end{equation}
where $E_{P}$ is the parameter related to the interaction between the conduction band and valence bands, $\Psi$ is the wave function of 2DEG along the growth axis $z$, and $E_{F}$ is the Fermi energy. $E_{\Gamma_{i}}(z)$ is defined as the band-edge energy of the $\Gamma_{iv}$ ($i=$7, 8) valence band at $z$.

\begin{figure}[!tp]
\includegraphics[width=2.8in,clip=]{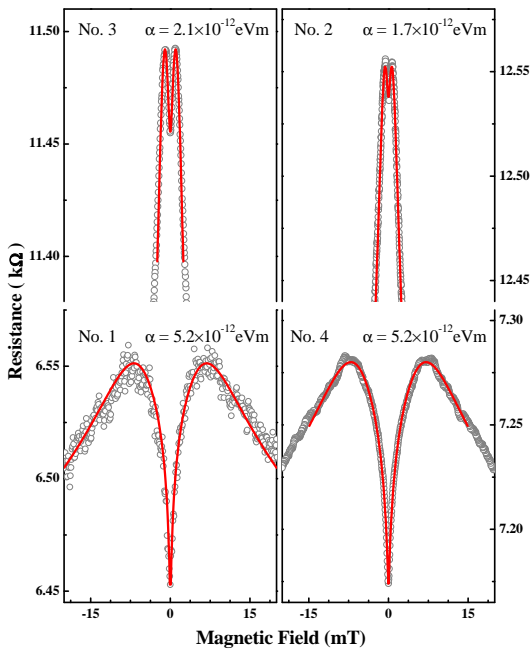}
\caption{\label{fig:RB}Longitudinal resistance ($R_{xx}$) versus magnetic field for the four samples at 0.3 K. The experimental results (circles), as well as calculated ones (solid curves), are compared with similar carrier densities in the same doping positions.  The gate-controlled carrier densities are, for the front-doping samples, $4.3\times 10^{11}$ cm$^{-2}$ (No. 1) vs. $4.5\times 10^{11}$ cm$^{-2}$ (No. 3), and, for the back-doping samples, $5.9\times 10^{11}$ cm$^{-2}$ (No. 2) vs. $6.0\times 10^{11}$ cm$^{-2}$ (No. 4). For samples with the front- (back-) doping, the SO coupling constant $\alpha$ is larger in No. 1 (No. 4), which has (does not have) the InP/In$_{0.53}$Ga$_{0.47}$As.}
\end{figure}

Figure~\ref{fig:RB} shows the selected WAL results for the four samples with similar carrier densities in the left and right panels. The dip in magnetoresistance is the signature of the SO interaction in 2DEG. The field at which the maximum resistance occurs corresponds to $H_{SO}$, and $H_{SO}$ is an indication of the strength of the SO interaction since $H_{SO}$ is proportional to $\alpha^{2}$. As clearly shown in the left panel of Fig.~\ref{fig:RB}, the SO interaction in No. 3 was much weaker than that in No. 1 for the front-doping condition. Since the difference in the carrier density was less than 5\%, it is possible that the InP/In$_{0.53}$Ga$_{0.47}$As interface that accounts for the enhancement of the $\alpha$ value in the front-doping case. For the samples with the back-doping condition, No. 2 and No. 4, the situation is reversed: a weaker SO interaction was observed in sample No. 2 which had an inserted InP. These observations are consistent with what the $\mathbf{k\cdot p}$ formalism [Eq.~(\ref{eq:a})] predicts as discussed below.

The way the doping position and the interface affects the SO interaction can be understood qualitatively from the coupling constant $\alpha$ expressed in the $\mathbf{k \cdot p}$ formalism. Contributions to Eq.~(\ref{eq:a}) can be split into two parts: (i) the field part ($\alpha_{f}$), which is related to the electric field within the QW and (ii) the interface part ($\alpha_{i}$), which is related to some band discontinuities in valence bands at hetero-interfaces. $\alpha_{f}$ is the expected value of the electric field in the active region with the band parameters as prefactors, $C_{f}= (E_{F}-E_{\Gamma_{7}})^{-2}-(E_{F}-E_{\Gamma_{8}})^{-2}$. Since the sign of $C_{f}$ is fixed for all materials, the sign of $\alpha_{f}$ is determined by the electric field, and therefore is affected by the doping position~\cite{Schapers98} and the gate voltage.~\cite{Schultz96,Nitta97,Lu98}

On the other hand, the interface-part contribution, either additive or subtractive to the field part, is more complicated due to the prefactors ($C_{ix}$) of the electron probabilities at interfaces: $\alpha_{i}=-(C_{iu}|\Psi_{u}|^{2}-C_{il}|\Psi_{l}|^{2}$), where $|\Psi_{x}|^{2}$ is electron probability at the interface $x=u$ (upper) or $l$ (lower). In the simplest case, \emph{i.e.} identical interfaces, the sign of $\alpha_{i}$ is determined by the difference of electron probabilities at interfaces, which is related to the electric field and eventually gives the subtractive effect to the field part.~\cite{Engles97,Schapers98} To have the additive contribution in $\alpha$ value, the interfaces should be different. $C_{ix}$, whose denominator is similar to $C_{f}$'s, is related to the offset energies of valence bands.~\cite{Engles97} These offset energies can influence the sign of $\alpha_{i}$. Due to the smaller $\Delta_{SO}$ in InP, the $\Gamma_{8v}$ band offset is larger than the $\Gamma_{7v}$ one at InP/In$_{0.53}$Ga$_{0.47}$As, which makes $C_{iu} > C_{il}$ and then leads to the negative $\alpha_{i}$ (see Fig.~\ref{fig:bands}). Therefore, $\alpha_{i}$ is additive to $\alpha_{f}$ when the InP layer is placed on the same side of doping position, like No.~1 where the sign of electric field is negative too; but it is subtractive in the opposite way (No.~2). Under the same doping position with similar $N_{S}$, the former enhances the overall $\alpha$ value [\emph{i.e.}, No.~1 showed a larger opening in $R_{xx}(H)$ than No.~3 did], while the latter reduces the overall value [\emph{i.e.}, No.~2 showed a smaller opening in $R_{xx}(H)$ than No.~4 did].

\begin{figure}[!h]
\includegraphics[width=2.6in,clip=]{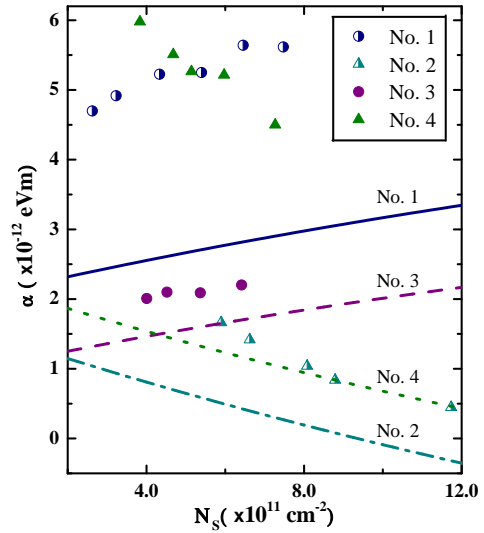}
\caption{\label{fig:aNs}Experimental results (symbols) and calculations (lines, labeled separately) of $\alpha$ versus $N_{s}$ for the four samples. For the front- (back-) doping samples, $\alpha(N_{S})$ shows the positive (negative) slope and the SO interaction is enhanced (reduced) due to the existence of the InP/In$_{0.53}$Ga$_{0.47}$As interface. Front-doping samples are No. 1 (InP) and No. 3, and back-doping ones are No. 2 (InP) and No. 4.}
\end{figure}

The above interpretation from the $\mathbf{k \cdot p}$ formalism can explain the results in Fig.~\ref{fig:RB} only qualitatively. Figure~\ref{fig:aNs} shows the dependences of the experimental $\alpha$ value (symbols) on carrier density $N_{S}$ for all samples, as compared with those from the calculations (curves). As expected, the sign of $d\alpha/dN_{s}$ is positive (negative) when the doping position is above (below) the well, as seen in Nos.~1 and 3 (Nos.~2 and 4). For the same doping condition (\emph{i.e.} the same sign of the field-part contribution), the overall $\alpha$ values were enhanced (reduced) in No. 1 (No. 2) relative to those in No. 3 (No. 4), where both field and interface contributions to the SO coupling were additive (subtractive). However, despite the slope ($d\alpha/dN_{s}$) and the interface effect meet our expectations qualitatively, the magnitudes of $\alpha$ values for all samples were significantly large.

To clarify the causes for this discrepancy, we need to examine both the calculation and the actual sample conditions in more details. One crucial point in the calculation is the knowledge of the precise potential profile. The band-structure profile, \emph{e.g.}, Fig.~\ref{fig:bands}, is normally obtained by solving Schr\"odinger-Poisson equations self-consistently, which requires the Fermi pinning energies as boundary conditions. These pinning energies, however, were not known in our samples: one located on the surface of our samples, and the other near the substrate/buffer layer interface. We have carefully designed samples and measurements to extract this information. But having the correct potential profiles did not significantly affect the calculation results. Another adjustment in the calculation would be to include the background impurity concentration,~\cite{Schapers98,Koga03b} which would shift the whole curve of $\alpha(N_{S})$ vertically. Had we included the background impurities to compensate for the big gap between experiments and calculations, the Fermi energy in some samples would have become higher than that of the conduction band in the carrier-supply layer. It is unlikely that we have such a situation for our samples.

Another possible cause, a more practical one, for the discrepancy between the measurements and calculations could be the qualities of the materials themselves, especially in the inserted InP layer and the interfaces. Cross-sectional transmission electron microscope (TEM) images of the layer structures clearly revealed that an unknown compound was formed in the In$_{0.52}$Al$_{0.48}$As/InP interface. This compound formation might have occurred in the InP/In$_{0.53}$Ga$_{0.47}$As interface as well, though it was not as obvious as that at the In$_{0.52}$Al$_{0.48}$As/InP interface because of the similar colorings between them. It is well known that InAsP islands reside in the In$_{0.53}$Ga$_{0.47}$As/InP interface,~\cite{Bohrer92} and the InP layer in our samples was intentionally placed above the QW to avoid this problem. However, we are not sure whether our InP/In$_{0.53}$Ga$_{0.47}$As interfaces exhibited the As-P exchange effect~\cite{Anan93} and tensile strain~\cite{McKay03} or not, as observed in other kinds of growth methods. A further analysis by TEM with an energy dispersive X-ray spectrometer indicated that the inserted ``InP'' layer partially contained Ga and As. Besides, the In$_{0.53}$Ga$_{0.47}$As well showed some inhomogeneousness in thickness. This could have had a significant effect on the calculation results, where only pure materials and clean interfaces were assumed. The strain effect in a QW structure may cause an anomalous spin-orbit effect.~\cite{Studenikin03a} However, the argument about InP/In$_{0.53}$Ga$_{0.47}$As does not apply in No. 4 that lacks an InP layer. The deviations of $\alpha$ values in No. 4 were larger than those in Nos. 1--3. To find out the mechanisms of this abnormal result is one of our future research topics.

To summarize, we have studied the interface effect on the Rashba SO interaction in In$_{0.52}$Al$_{0.48}$As/In$_{0.53}$Ga$_{0.47}$As QWs by a weak antilocalization analysis. Introducing an InP layer above the QW can strengthen or weaken the SO interaction by incorporating the effect of the front or back doping position, respectively. According to the doping position, $d\alpha/dN_{S}$ can be either positive (front-doped) or negative (back-doped). These phenomena can be understood from the $\mathbf{k\cdot p}$ formalism of the SO coupling constant $\alpha$. Furthermore, providing attainable growth conditions, one can tailor the layer structure for a maximal or minimal interface effect on the $\alpha$ value. Besides the observations as predicted, there is some discrepancy in the magnitudes between the experimental and calculated $\alpha$ values. This discrepancy can be attributed to the actual conditions of the interfaces and materials in our samples.


\begin{thebibliography}{29}
\bibitem[{TEM()}]{TEM}
\bibinfo{note}{Not the actual thickness revealed by TEM images (discussed in
  the end).}
\end{thebibliography}
\end{document}